\documentclass[twocolumn,superscriptaddress,amsfont,amssymb,amsmath, showpacs,balancelastpage,longbibliography,nofootinbib]{revtex4-1}
\usepackage{CJKutf8}
\usepackage[normalem]{ulem}
\usepackage{amsthm}

\usepackage[utf8]{inputenc}
\usepackage{amsmath,amssymb,amsfonts,stmaryrd}
\usepackage{physics}
\usepackage{dsfont}
\usepackage{graphicx}
\usepackage{xcolor}
\usepackage{graphicx}
\usepackage{cancel}
\usepackage{natbib}
\usepackage{color}
\usepackage{tikz}
\usepackage{mathrsfs}
\usepackage{relsize}
\usepackage{multirow}
\usepackage[linktocpage]{hyperref}
\usepackage[all]{xy}
\hypersetup{colorlinks=true,citecolor=blue,linkcolor=blue, urlcolor=blue, breaklinks=true}

\usepackage{float}

\usepackage{bm} 
\usepackage{subfigure} 
\usepackage{environ}
\usepackage{url}
\usepackage[margin=0.75in]{geometry}
\usepackage{ulem}

\usepackage{mathtools}

\usepackage{amsmath}

\newcommand{\Z}{{\mathbb Z}}

\NewEnviron{eqs}{%
\begin{equation}\begin{split}
    \BODY
\end{split}\end{equation}}

\definecolor{dkgreen}{rgb}{0,0.5,0}

\theoremstyle{definition}

\theoremstyle{remark}

\begin{document}

\begin{CJK*}{UTF8}{bsmi}

\title{(2+1)D topological phases with RT symmetry: \\ 
many-body invariant, classification, and higher order edge modes}

\author{Ryohei Kobayashi}

\affiliation{Department of Physics and Joint Quantum Institute, University of Maryland, College Park, Maryland 20742, USA}
\affiliation{Condensed Matter Theory Center, University of Maryland, College Park, Maryland 20742, USA}

\author{Yuxuan Zhang}
\affiliation{Department of Physics and Joint Quantum Institute, University of Maryland,
College Park, Maryland 20742, USA}
\affiliation{Condensed Matter Theory Center, University of Maryland, College Park, Maryland 20742, USA}

\author{Yan-Qi Wang}
\affiliation{Department of Physics and Joint Quantum Institute, University of Maryland,
College Park, Maryland 20742, USA}

\author{Maissam Barkeshli}
\affiliation{Department of Physics and Joint Quantum Institute, University of Maryland,
College Park, Maryland 20742, USA}
\affiliation{Condensed Matter Theory Center, University of Maryland, College Park, Maryland 20742, USA}

\begin{abstract}
It is common in condensed matter systems for reflection ($R$) and time-reversal ($T$) symmetry to both be broken while the combination $RT$ is preserved. In this paper we study invariants that arise due to $RT$ symmetry. We consider many-body systems of interacting fermions with fermionic symmetry groups $G_f = \mathbb{Z}_2^f \times \mathbb{Z}_2^{RT}$, $U(1)^f \rtimes \mathbb{Z}_2^{RT}$, and $U(1)^f \times \mathbb{Z}_2^{RT}$. We show that (2+1)D invertible fermionic topological phases with these symmetries have a $\mathbb{Z} \times \mathbb{Z}_8$, $\mathbb{Z}^2 \times \mathbb{Z}_2$, and $\mathbb{Z}^2 \times \mathbb{Z}_4$ classification, respectively, which we compute using the framework of $G$-crossed braided tensor categories. We provide a many-body $RT$ invariant in terms of a tripartite entanglement measure, and which we show can be understood using an edge conformal field theory computation in terms of vertex states. For $G_f = U(1)^f \rtimes \mathbb{Z}_2^{RT}$, which applies to charged fermions in a magnetic field, the non-trivial value of the $\mathbb{Z}_2$ invariant requires strong interactions. For symmetry-preserving boundaries, the phases are distinguished by zero modes at the intersection of the reflection axis and the boundary. Additional invariants arise in the presence of translation or rotation symmetry. 
\end{abstract}

\maketitle

\end{CJK*}

{\it Introduction. }A significant direction in condensed matter physics is to develop a comprehensive understanding of topological invariants that can distinguish quantum phases of matter \cite{thouless1982,wen04,bernevig2013topological,hasan2010,qi2010RMP,nayak2008,Senthil2015SPT,barkeshli2019,zeng2019quantum}. Recently there have been breakthroughs for strongly interacting systems of fermions \cite{Barkeshli_2022_PRB_Classification,Wang_2020_PRX_Construction,aasen2021characterization,Bulmash2022fermionic} with physically relevant crystalline symmetries \cite{song2017topological,Huang2017,shiozaki2017invt,manjunath2021cgt,manjunath2020FQH,zhang2022fractional,zhang2022pol,zhang2022real,Manjunath_2023_PRB_Nonperturbative,zhang2023complete,manjunath2024characterization,herzog2024interacting}. 
In this paper, we focus on a commonly occurring symmetry in condensed matter systems, the combination of reflection and time-reversal symmetry, denoted $RT$, and which generates an order-2 group $\mathbb{Z}_2^{RT}$. A common scenario, as in quantum Hall systems, chiral superconductors and magnets, is for reflection ($R$) and time-reversal ($T$) to be separately broken, either by an applied magnetic field or spontaneously, while the combination $RT$ remains a symmetry. This raises the question of whether new topological phases appear with $RT$ symmetry and what the corresponding invariants are. The case where $R$ and $T$ are individually symmetries of the interacting problem has been studied to some extent \cite{yao2013interaction}.

In this paper, we consider systems with the following fermionic symmetry groups: $G_f = \mathbb{Z}_2^f \times \mathbb{Z}_2^{RT}$, $U(1)^f \rtimes \mathbb{Z}_2^{RT}$, and $U(1)^f \times \mathbb{Z}_2^{RT}$. Here $\mathbb{Z}_2^f$ refers to the order-2 group generated by fermion parity, $(-1)^F$, and $U(1)^f$ refers to the group $U(1)$ where the $\pi$ rotation is fermion parity. $G_f = \mathbb{Z}_2^f \times \mathbb{Z}_2^{RT}$ is relevant for describing mean-field Bogoliubov-de-Gennes (BdG) Hamiltonians for chiral superconductors; $G_f = U(1)^f \rtimes \mathbb{Z}_2^{RT}$ is relevant for describing charged fermions in a magnetic field; $G_f = U(1)^f \times \mathbb{Z}_2^{RT}$ is relevant when the $U(1)^f$ describes spin rotation symmetry. We discuss the effect of lattice translations and rotations at the end of the paper.

\begin{table}[t]
    \centering
    \begin{tabular}{c|c|c|c}
       $G_f$ & Classification  & Labels & $I_{RT}$ \\
       \hline 
       $\mathbb{Z}_2^f \times \mathbb{Z}_2^{RT}$ & $\mathbb{Z} \times \mathbb{Z}_8$ & ($2c_-, \nu$) & $e^{2\pi i (c_- + \nu)/8}$ \\
       $U(1)^f \rtimes \mathbb{Z}_2^{RT}$ & $\mathbb{Z}^2 \times \mathbb{Z}_2$ & $(c_-, C, \nu)$ & $e^{2\pi i (c_- + 4\nu)/8}$\\
       $U(1)^f \times \mathbb{Z}_2^{RT}$ & $\mathbb{Z}^2 \times \mathbb{Z}_4$ &$(c_-, C, \nu)$ & $e^{2\pi i (c_- + 2\nu)/8}$\\
\hline       
    \end{tabular}
    \caption{Classification of (2+1)D invertible fermionic topological phases with $G_f$ symmetry, labeled by the chiral central charge $c_-$, Chern number $C$, and an integer index $\nu$.}
    \label{tab:Classif_Summary}
\end{table}

\begin{figure}[t]
\centering 
\includegraphics[width=0.85\columnwidth]{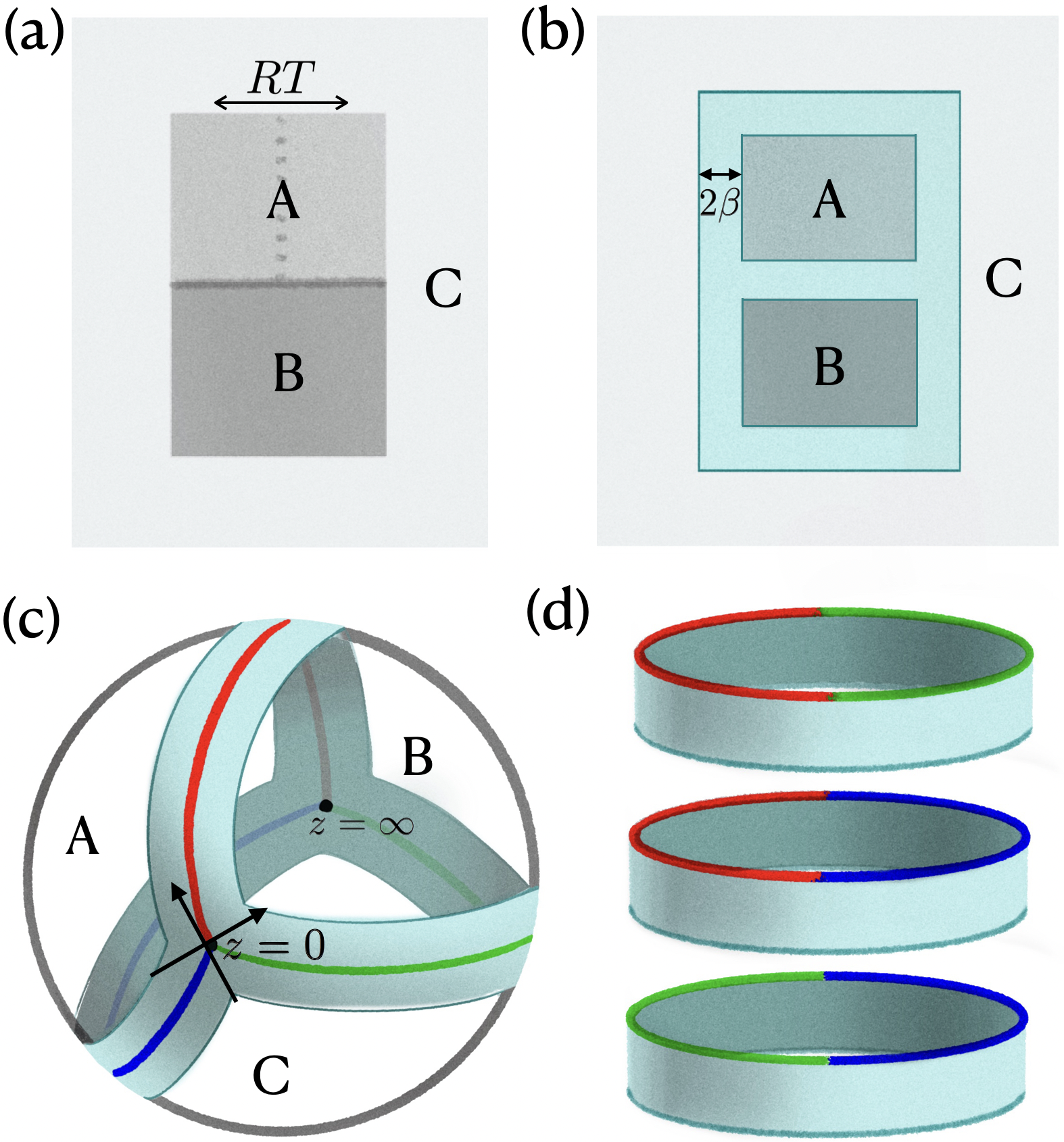}
\caption{(a) Placement of Region A, B, C. Partial $RT$ only acts on A. (b) The teal region represent the 3-vertex state $\ket{\psi_{\text{ABC}}}$. (c) $\ket{\psi_{\text{ABC}}}$ is topologically equivalent to a 3-punctured sphere. (d) $\ket{\psi_{\text{ABC}}}$ after performing the conformal transformation $z\rightarrow z^{3/2}$ where boundaries of the same color are identified.}
\label{fig:partialRT_schematic}
\end{figure}

{\it Definition of the invariant. } We begin by presenting a many-body invariant arising from $\mathbb{Z}_2^{RT}$ symmetry, and which can be extracted from a single bulk ground state wave function, adding to previous work extracting invariants from single bulk wave functions 
\cite{levin2006,kitaev2006topological,shiozaki2017invt,dehghani2021,cian2021,cian2022extracting,Kim2022ccc,fan2022,zhang2023complete}. Following ideas of implementing partial symmetry operations  to extract invariants \cite{Qi2012momentumpolarization,FQHEDMRG,Zaletel2014bosonicSPT,shiozaki2017invt, Shiozaki2018antiunitary, kobayashi2019,You2020hoe,zhang2023complete,herzog2024interacting, kobayashi2024hcc}, we wish to define a ``partial" $RT$ transformation, whose ground state expectation value defines our invariant. Consider a (2+1)D gapped system on flat space. The reflection acts on space as $ R : (x,y)\to(-x,y)$ while time-reversal $T = U\mathcal K$, with ${\mathcal K}$ denoting complex conjugation and $U$ a unitary operator. $U f_{i} U^\dagger =f_{j} \mathcal{U}_{ij}$, where $i,j$ index sites and flavors of the fermion annihilation operator $f$. 
$U$ only acts on the on-site degrees of freedom, so $\mathcal{U}_{ij}=0$ if $i$ and $j$ are at different positions.  We assume that the entire system is invariant under the combined $RT$ symmetry operation, with $(RT)^2=1$. We take a tripartition of the system into subsystems A, B, C, such that A and B are adjacent disks that respect the reflection symmetry, see Figure \ref{fig:partialRT_schematic} (a). 
Consider the reduced density matrix $\rho_{\mathrm{AB}} ={\rm Tr}_{\rm C}(\ket{\Psi}\bra{\Psi})$, where $\ket{\Psi}$ is the ground state wave function of the entire system. We define:
\begin{equation}\label{Eq_2d}
	 Z_{RT} := {\rm Tr}_{\mathrm{AB}} ( R_\mathrm{A} U_\mathrm{A} \rho_{\mathrm{AB}}^{T_\mathrm{A}} U_\mathrm{A}^\dagger R_\mathrm{A}^\dagger \rho_{\mathrm{AB}})
\end{equation}
and $I_{ RT} := Z_{ RT}/|Z_{RT}|$. 
$R_{\rm A}$ is the reflection operator restricted to region A, meaning that it acts as a reflection within A and as the identity outside of A. $\rho_{\rm AB}^{T_{\rm A}}$ denotes the fermionic partial transpose \cite{ShapourianPRL2017}, which allows us to define a partial time-reversal operation. The many-body ground state can be decomposed into the coherent state basis $\ket{\{ \xi_j \}}:=e^{-\sum_j\xi_jf_j^{\dagger}}\ket{0}$, $\bra{\{ \bar \chi_j \}}:=\bra{0}e^{-\sum_jf_j\bar \chi_j}$ where $\chi$ and $\xi$ are Grassmann variables.
The density matrix is expanded as:
\begin{equation}
    \rho_{\rm AB} = \int d[\bar \xi, \xi] d[\bar \chi, \chi]  \rho_{\rm AB}(\{ \bar \xi_j \}; \{ \chi_j\} ) \ket{\{ \xi_j \}_{j }}\bra{\{ \bar \chi_j \}_{j } },
\end{equation}
where $d[\bar \xi, \xi] := \prod_j d\bar \xi_j d \xi e^{- \sum_j \bar \xi_j \xi_j}$, and $\rho_{\rm AB}(\{ \bar \xi_j \};\{\chi_j\}) := \bra{\{ \bar \xi_j \}} \rho_{\rm AB} \ket{\{ \chi_j \}}$. The combined position and flavor index $j$ runs through all the sites in ${\rm A} \cup {\rm B}$. The partial time reversal of $\rho_{AB}$ is:
\begin{equation}
\begin{aligned}
    &U_{\rm A} \rho_{\rm AB}^{T_{\rm A}} U_{\rm A}^\dagger =  \int [d\bar \xi, \xi] d[\bar \chi, \chi]\rho_{\rm AB}(\{ \bar \xi_j \}; \{ \chi_j \}) \\
    &\bigl|\{ i \mathcal{U}_{ji} \bar \chi_j\}_{j \in {\rm A}}, \{ \xi_j\}_{j \in {\rm B}} \bigl>  \bigl<\{ i\xi_j\mathcal{U}_{ji}^{\dagger} \}_{j \in {\rm A}}, \{ \bar \chi_j \}_{j \in {\rm B}} \bigl|.
\end{aligned}
\end{equation}
We will see that $I_{RT}$ is a quantized invariant, up to corrections exponentially small in the size of $AB$, with the quantization shown in Table \ref{tab:Classif_Summary}.

To motivate Eq. \ref{Eq_2d}, note that at the reflection axis $x=0$, $RT$ effectively reduces to an internal anti-unitary symmetry. Therefore, for any given state, we can consider stacking at $x = 0$ a (1+1)D invertible fermionic state with $\mathbb{Z}_2^{T}$ symmetry. When $G_f = \mathbb{Z}_2^f \times \mathbb{Z}_2^{RT}$, the (1+1)D state is in Class BDI \cite{Altland1997,Kitaev2009periodic,schnyder2008}, which is known to have a $\mathbb{Z}_8$ classification \cite{fidkowski2011}. Viewing the system as an effectively one-dimensional system along the $y$ direction, our many-body invariant reduces to the many-body invariant presented in \cite{Shiozaki2018antiunitary} for (1+1)D topological phases in class BDI. Similarly, for $G_f = U(1)^f \rtimes \mathbb{Z}_2^{RT}$ and $U(1)^f \times \mathbb{Z}_2^{RT}$ the corresponding (1+1)D states are in symmetry classes AI and AIII, respectively. 

\begin{figure}[t]
\centering 
\includegraphics[width=0.95\columnwidth]{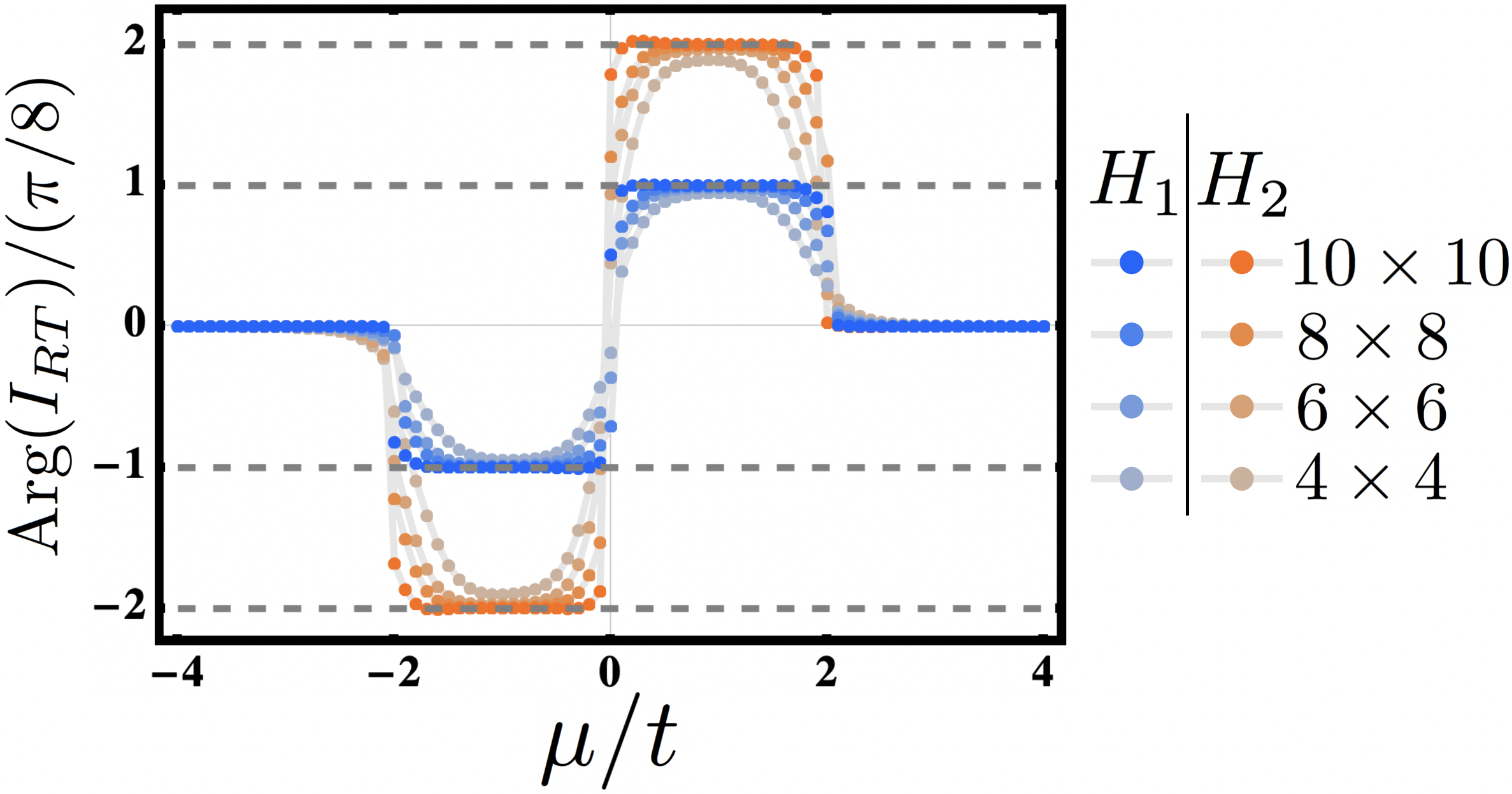}
\caption{\label{Fig_RT_Different_Reflection} The phase of $RT$ invariant $I_{RT}$ for the $H_{1}$ (single layer) and $H_2$ (bilayer) in units of $\pi/8$. The numerics are done on a $16 \times 16$ square lattice with periodic boundary condition. Different  saturation of the data points represent different system sizes of the $A\cup B$ region.}
\end{figure}

\begin{figure*}[t]
\centering 
\includegraphics[width=2\columnwidth]{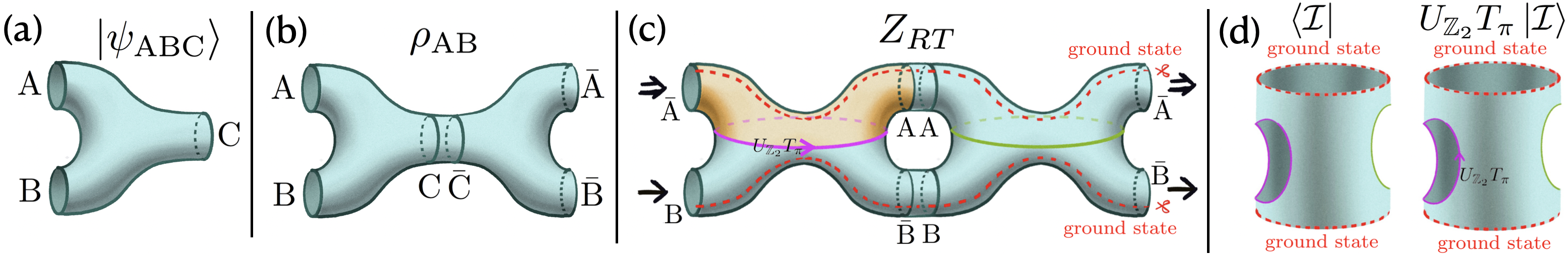}
\caption{(a) The 3-vertex state $\ket{\psi_{\text{ABC}}}$ is topologically equivalent to a pair of pants. (b) 
$\rho_{\text{AB}}= \text{Tr}_{\text{C}}(\ket{\psi_{\text{ABC}}}\bra{\psi_{\text{ABC}}})$ is constructed by gluing two 3-vertex states with opposite orientations along $C$. (c) $Z_{RT}$ is constructed by gluing 2 $\rho_{\text{AB}}$ with the top half of the first $\rho_{\text{AB}}$ (tinted yellow) twisted by $U_{\mathbb{Z}_2}T_{\pi}$. We could cut at the red dashed line by inserting ground states, and fictitiously cut through the magenta and green circle, which result in (d): two copies of Ishibashi state $\ket{\mathcal{I}}$, with one of them twisted by $U_{\mathbb{Z}_2}T_{\pi}$.}
\label{fig:ishibashi}
\end{figure*}

{\it Numerical results. } We confirm the above statements empirically via numerical calculations of $I_{RT}$ for multiple layers of chiral p-wave superconductors. A single layer of chiral p-wave  superconductor has BdG Hamiltonian
\begin{equation}\label{Eq_Single_Layer_SC}
    H_s \!=\! \!-\!\sum_{\vec i} \mu_{\vec i} f^\dagger_{\vec i,s} f_{\vec i,s} \!-\! \frac{1}{2} \sum_{\langle \vec i, \vec j \rangle} [t f^\dagger_{\vec i,s} f_{\vec j,s}+ \Delta^{(s)}_{\vec i, \vec j} f^\dagger_{\vec i,s} f^\dagger_{\vec j,s}+ {\rm H.c.}].
\end{equation}
Here $s = \pm 1$ denotes its chirality, $\vec i = (x_i,y_i)$ denotes the coordinate of $i-$th site on a square lattice expanded by the unit vector $(\hat e_x, \hat e_y)$, and the summation $\langle \rangle$ runs over nearest neighbors. The pair order parameter $\Delta^{(s)}_{\vec i,\vec i+\hat e_x} = is\Delta$, $\Delta^{(s)}_{\vec i, \vec i+ \hat e_y} = \Delta$, with $\Delta, t > 0$. We have: ${T} f_{\vec j,s} { T}^{-1} = f_{\vec j,s}$, ${T} i {T}^{-1} = -i$, ${R} f_{(i_x,i_y),s} {R}^{-1} = f_{(-i_x,i_y),s}$. We include chemical potential disorder $\mu_{\vec i} = \mu + \delta_{\vec i}$, $|\delta_{\vec i}| < |\mu|$ in such a way that the Hamiltonian \eqref{Eq_Single_Layer_SC} breaks translation symmetry in both directions but respects ${R} {T}$ symmetry. We consider the following two cases: (1) a single layer chiral superconductor $H_1 = H_+$, (2) two layers of superconductors with opposite chiralities $H_2 = H_+ + H_-$. In both cases, the Hamiltonians $H_{1,2}$ have the symmetry $\Z_2^f \times \Z_2^{RT}$. 
We compute $I_{RT}$ numerically, with the results shown in Figure~\ref{Fig_RT_Different_Reflection}. These results agree with the proposed formula $I_{RT}=e^{2\pi i(c_- + \nu)/8}$ in Table \ref{tab:Classif_Summary}. Consider the case for $0<\mu<2t$. For $H_1$, the phase of $I_{RT}$ is quantized to $2\pi/16$, which agrees with $c_- = 1/2, \nu = 0 \text{ mod }8$, and for $H_2$, the phase of $I_{RT}$ is quantized to $2\pi/8$, which agrees with $c_-=0,\nu=1\text{ mod }8$. Free fermion models for $U(1)^f \times \mathbb{Z}_2^{RT}$ can be obtained by taking two copies of the above systems.

{\it Computation of invariant from edge CFT. } A central result of this paper is that we can derive the properties of $Z_{RT}$, assuming the equivalence \cite{Haldane2008entanglement} between the entanglement spectrum and an idealized CFT description of the edge theory. For simplicity, we assume that the edge CFT is chiral.~\footnote{In the case of non-chiral edge CFT, it breaks into tensor product of chiral and anti-chiral modes, and one can compute each part separately.}
Below we sketch the calculation; additional details are presented in Appendix \ref{app:cft}. 
We employ the ``cut-and-glue'' argument~\cite{qi2012entanglement}, where we first physically cut the whole system into the tripartition, which gives rise to edge theories along the boundaries of each subsystem. We then ``glue'' the cut back by turning on the interaction connecting the subsystems along the interface. This gluing interaction is regarded as the perturbation gapping out the counter-propagating CFT at the cut, and thought to correspond to the conformally invariant boundary condition of the CFT~\cite{calabrese2007quench}. This perspective allows us to identify the resulting state after gluing as the boundary state of the CFT.

If we had a bipartition of the 2d system instead of the tripartition, we would have been able to explicitly represent the boundary state of edge CFT at the cut using the Ishibashi state~\cite{Ishibashi}. But given we consider the tripartition of the system, there is no known explicit form for the boundary state. 
However, one can still express the CFT state at the tripartition cut in terms of the path integral of the CFT~\cite{liu2023multipartite}, where we take the 2D spacetime for the (1+1)D edge CFT to be a network of thin slabs obtained by fattening the tripartition cut by a width $2\beta$, see Figure \ref{fig:partialRT_schematic} (b). The parameter $\beta$ is regarded as a regulator introduced to normalize the boundary state, and $\beta\to 0$ is understood. 

Suppose that the 2d space for the bulk wavefunction is located on a sphere and A, B, C are the tripartition of a sphere. The path integral for the CFT boundary state on the tripartition cut $\ket{\psi_{\mathrm{ABC}}}$ is then carried out on an open surface topologically equivalent to a 2D sphere with three punctures, see Figure \ref{fig:partialRT_schematic} (c). By performing the conformal transformation $z\to z^{3/2}$ on the punctured sphere, it is transformed to an open surface obtained by gluing three annuli with width $\beta$ at their boundary, see Figure \ref{fig:partialRT_schematic} (d). The resulting boundary state after the conformal transformation is referred to as the 3-vertex state. From now, when we refer to a 3-punctured sphere for $\ket{\psi_{\mathrm{ABC}}}$ we consider the one after the transformation $z\to z^{3/2}$.
$\rho_{\mathrm{AB}}$ is then represented by a path integral on a surface obtained by gluing a pair of punctured spheres $\ket{\psi_{\mathrm{ABC}}}$ along the boundary circle which corresponds to the Hilbert space of the C subsystem. Gluing along the C circle amounts to tracing out the Hilbert space for C subsystem. The resulting surface is topologically equivalent to a sphere with four punctures, see Figure \ref{fig:ishibashi} (b). 

Let us now express the partial $RT$ transformation at the level of the CFT path integral. The partial $RT$ acts on the Hilbert space for the A circle, and has the effect of reflecting the spatial A circle as well as exchanging the bra and ket of $\rho_{\mathrm{AB}}$ due to the partial transpose acting on A. These effects are simultaneously achieved by a half Dehn twist inserted at the equator of the 4-punctured sphere, where the A circles are located at the northern hemisphere (see Figure \ref{fig:ishibashi} (c)). 
Let us recall that the $\pi$ rotation in the Euclidean spacetime reversing the time direction is identified as the $CRT$ symmetry in Lorentz invariant theory~\cite{Hason_2020}. We then postulate that our $RT$ symmetry is the combined operation of $CRT$ and internal $\Z_2$ symmetry. 
This identification then allows us to express the partial $RT$ in terms of the half Dehn twist associated with an insertion of the $\Z_2$ symmetry defect along the equator of the 4-punctured sphere.

The invariant $Z_{RT}$ then corresponds to a path integral on a closed Riemann surface, obtained by a pair of 4-punctured spheres where the A, B boundaries are connected up, and one of the punctured sphere is inserted with the half Dehn twist associated with the $\Z_2$ symmetry defect. 
This resulting Riemann surface is a genus 3 surface with a defect insertion, and it is generally formidable to compute the CFT partition function on a genus 3 surface. However, in the limit of $\beta\to 0$ where the width of the annulus is thin, one can greatly simplify the topology and work out the computation.

When $\beta\to 0$, if we regard the direction along the annulus as a fictitious time direction, the Hilbert space living at the interval with length $2\beta$ is in the low temperature limit, and effectively realizes its ground state. So one can insert the projector onto the ground state along a cut (see Appendix \ref{app:cft} for further details). In the 4-punctured sphere for $\rho_{\mathrm{AB}}$, we insert the projector along a line connecting the pair of A circles and a line connecting the pair of B circles. For $Z_{RT}$ this amounts to cutting along the (red) dashed line in Figure \ref{fig:ishibashi}(c) and setting the boundary condition to be the ground state.  

Now, the invariant $Z_{RT}$ is the path integral on a surface shown in Figure \ref{fig:ishibashi} (c) with the ground state projector inserted along the red dashed lines. To evaluate the phase of this path integral, we cut the surface for $Z_{RT}$ in Figure \ref{fig:ishibashi} (c) along the purple and green curves, which results in two disconnected open surfaces as shown in Figure \ref{fig:ishibashi} (d). Each open surface shown in Figure \ref{fig:ishibashi} (d) is a cylinder with the circumference $8\beta$, and the ends of the cylinder are set to the ground state boundary condition. The pair of holes (purple and green) on each cylinder support a state of the CFT. As explained in the Appendix \ref{app:cft}, this state is the Ishibashi state $\ket{\mathcal{I}}$ in the limit $\beta\to 0$. 

Finally, the invariant $Z_{RT}$ is regarded as the expectation value in the Ishibashi state of the half Dehn twist $T_\pi$ followed by the $\Z_2$ symmetry generator $U_{\Z_2}$ acting on a single hole boundary (purple circle, which we can take to be the holomorphic part of the Hilbert space)
\begin{align}
    Z_{RT} = \bra{\mathcal{I}} U_{\Z_2} T_\pi \ket{\mathcal{I}}. 
\end{align}
The insertion of the symmetry operator reflects the defect insertion at one of the equators in Figure \ref{fig:ishibashi}(c). This Ishibashi state $\ket{\mathcal{I}}$ is normalized such that the norm of the state is given by the 2nd Renyi entropy in the limit $\beta \rightarrow 0$, 
$\mathrm{Tr}(\rho_{\mathrm{AB}}^2) = \bra{\mathcal{I}} \ket{\mathcal{I}}$,
which has been computed in~\cite{liu2023multipartite}.

The phase of $Z_{RT}$ can be evaluated by identifying the Ishibashi state as the path integral on a single annuli with thin width $\beta'\to 0$. This $\beta'$ is regarded as another way of regularizing the Ishibashi state $\ket{\mathcal{I}}$ at $\beta\to 0$ defined above, and we assume that the result of the invariant $I_{RT}$ does not depend on the choice of the regularization. 
Then, $Z_{RT}$ is regarded as connecting two annuli along their boundaries with a single insertion of the defect $ U_{\Z_2} T_\pi$. This is the torus partition function with the defect insertions, which can be expressed in terms of the chiral Virasoro character,
\begin{align}
    {Z}_{RT} \propto e^{\frac{2\pi i}{48}c_-}\chi_1\left(\tau= i\beta' + \frac{1}{2},(0,1), (\mathrm{AP,AP})\right),
\end{align}
where $\propto$ denotes up to overall (real) normalization, and $(0,1)$ denotes $\Z_2$ twisted sector on a torus. (AP, AP) represents the anti-periodic spin structure along each cycle.
The modular parameter $i\beta'+1/2$ reflects that the half Dehn twist along the spatial cycle is inserted.
 Explicit computations of this CFT character is carried out in Appendix \ref{app:cft}.
For instance, fermionic invertible phases with unitary $\Z_2$ symmetry in (2+1)D are labeled by $(\nu, 2c_-)\in \Z_8\times \Z$ \cite{gu2014}. $I_{RT}$ then evaluates to
\begin{align}
    I_{RT} = e^{\frac{2\pi i}{8}\nu }e^{\frac{2\pi i}{16}2c_-}.
\end{align}
Similar calculations for the case $U(1)^f \rtimes \mathbb{Z}_2$ and $U(1)^f \times \mathbb{Z}_2$ yield the results summarized in Table \ref{tab:Classif_Summary}.

{\it Classification.} The classification of phases summarized in Table \ref{tab:Classif_Summary} can be derived in two independent ways, exhibiting a remarkable agreement. Following the arguments in \cite{HuangPRB2017}, given two states we can apply a unitary circuit in any region $\mathcal{R}$ to locally transform one state to another; the circuit will be $RT$ symmetric if we apply an appropriate circuit to the reflected region $R(\mathcal{R})$. In this way, we can transform the two states to each other everywhere except for the reflection axis. Thus we expect that the classification of phases with $\mathbb{Z}_2^{RT}$ symmetry can be obtained by decorating the reflection axis with (1+1)D invertible states. The classification of (1+1)D fermionic invertible phases in class BDI, AI, and AIII are known to be $\mathbb{Z}_8$, $\mathbb{Z}_2$, and $\mathbb{Z}_4$, respectively \cite{fidkowski2011}. Combining this with the known invariants $c_-$ and Chern number $C$ gives Table ~\ref{tab:Classif_Summary}. 

For the case $G_f = {\rm U}(1)^f \rtimes \Z_2^{RT}$ the $\nu\in\Z_2$ index is {\it interaction-enabled}, meaning that it can only emerge in an interacting fermionic system. This is because (1+1)D strongly interacting invertible phases in Class AI have a $\mathbb{Z}_2$ classification, as compared with a trivial free fermion classification \cite{Kitaev2009periodic}. If a free fermion realization existed for our non-trivial $\nu$ index, then we could view the (2+1)D system as a (1+1)D system where the $x$ direction is thought of as an internal index, and we would arrive at a free fermion realization of the non-trivial Class AI phase, which is a contradiction.

Remarkably, we can independently derive the same classification more formally using the techniques developed in \cite{Barkeshli_2022_PRB_Classification,aasen2021characterization,Manjunath_2023_PRB_Nonperturbative}. To do this, we first use the \it fermionic crystalline equivalence principle (fCEP)\rm, which asserts that the classification of fermionic topological phases with a spatial symmetry $G_f$ is equivalent to one with an internal symmetry $G_f^{\text{eff}}$ \cite{Thorngren2018,Else2019,debray2021invertible,zhang2022real,Manjunath_2023_PRB_Nonperturbative}. Ref.~\onlinecite{Manjunath_2023_PRB_Nonperturbative} presented an algebraic formula for relating $G_f$ to $G_f^{\text{eff}}$. In general the fCEP is a conjecture which has received substantial empirical verification; the results here provide additional evidence for its validity. As reviewed in Appendix \ref{app:classification}, under fCEP one can replace the $\mathbb{Z}_2^{RT}$ symmetry with a unitary internal $\mathbb{Z}_2$ symmetry. Thus, we proceed by classifying fermionic invertible phases with symmetries $\mathbb{Z}_2^f \times \mathbb{Z}_2$, $U(1)^f \rtimes \mathbb{Z}_2 \simeq O(2)^f$, $U(1)^f \times \mathbb{Z}_2$. 

In general, invertible fermionic topological phases with internal symmetries can be systematically characterized and classified using an algebraic framework of $G$-crossed braided tensor categories \cite{barkeshli2019,Barkeshli_2022_PRB_Classification,aasen2021characterization}. Such phases are characterized by a set of data, $(c_-, n_1, n_2, \nu_3)$, where $n_1 \in C^1(G_b, \mathbb{Z}_2)$, $n_2 \in C^2(G_b, \mathbb{Z}_2)$, and $\nu_3 \in C^3(G_b, \mathbb{R}/\mathbb{Z})$. 
These data are subject to a set of consistency conditions, $dn_1 = 0$, $dn_2 = \mathcal{O}_3$, and $d \nu_3 = \mathcal{O}_4$. The expressions for $\mathcal{O}_3$ and $\mathcal{O}_4$, together with equivalence relations on the data $(n_1, n_2, \nu_3)$ are given in \cite{Barkeshli_2022_PRB_Classification} and summarized in Appendix \ref{app:classification}. By solving these consistency conditions and equivalence relations, we can derive the classification given in Table~\ref{tab:Classif_Summary}, as detailed in Appendix \ref{app:classification}.

\it Higher order edge modes\rm. If $c_- = C = 0$, the boundary does not generically have dispersing edge modes with the symmetries in Table \ref{tab:Classif_Summary}, unless additional symmetries are present. If the boundary respects the $RT$ symmetry, there will be an edge zero mode at the intersection point between the reflection axis and the boundary of the system. Thus such phases are analogous to ``higher order topological phases" \cite{Benalcazar2019HOTI} except the zero modes are not at corners. In the case of the $\nu =1$ phase with $G_f = \mathbb{Z}_2^f \times \mathbb{Z}_2^{RT}$, the intersection point localizes a Majorana zero mode. We can understand this in more detail for the model Hamiltonian $H_2$ introduced above. The low energy edge theory consists of counterpropagating Majorana fermions. A mass term $m(x)$ gaps the modes, but $RT$ symmetry requires $m(x) = -m(-x)$~\cite{wang2019boundary}. The domain wall in the mass leads to a Majorana zero mode at $x = 0$ \cite{read2000}. 

\it Including translation and rotation symmetry\rm. In the presence of translation or rotation symmetry, the system acquires inequivalent reflection axes, and we can decorate each inequivalent reflection axis with a (1+1)D invertible phase. Therefore, we expect to get additional invariants, one for each inequivalent reflection axis. For example, for a square lattice, we have four inequivalent reflection axes, one of which is a glide reflection axis, and so we expect four inequivalent $RT$ invariants. We leave a systematic classification for future work. 

\textit{Acknowledgements --}
RK thanks Yuya Kusuki and Yuhan Liu for useful discussions, and Shinsei Ryu for comments on a draft. YQW thanks Chunxiao Liu and Naren Manjunath for useful discussions. This work is supported by NSF DMR-2345644 and by an NSF CAREER grant (DMR- 1753240), the Laboratory for Physical Sciences through the Condensed Matter Theory Center, National Science Foundation QLCI grant OMA-2120757, and by the JQI theory postdoctoral fellowship (RK and YQW) at UMD.

\bibliography{bibliography}

\onecolumngrid

\vspace{0.3cm}

\begin{center}
\Large{\bf Supplemental Materials}
\end{center}
\onecolumngrid

\section{CFT computation of the invariant from vertex states}
\label{app:cft}
Here we provide the additional details for the computation of the invariant $Z_{RT}$ based on the chiral edge CFT. Suppose that the bulk 2d system is a sphere and A, B, C are its tripartitions. As outlined in the main text, the state $\ket{\Psi_{\mathrm{ABC}}}$ is represented as the path integral of chiral edge CFT on a network of thin slabs with width $2\beta$. This is topologically equivalent to a sphere with three punctures, where the cuts at the A, B, C subsystems are identified as punctures (see Figure \ref{fig:partialRT_schematic} (c) in the main text). By regarding the punctured sphere as a Riemann surface (with origin $z=0$ at the Y-junction) and performing the conformal transformation $z\to z^{3/2}$, the surface is transformed to three annuli glued along one boundary together, see Figure \ref{fig:partialRT_schematic} (d) in the main text. 

\begin{figure}[t]
\centering 
\includegraphics[width=1.0\columnwidth]{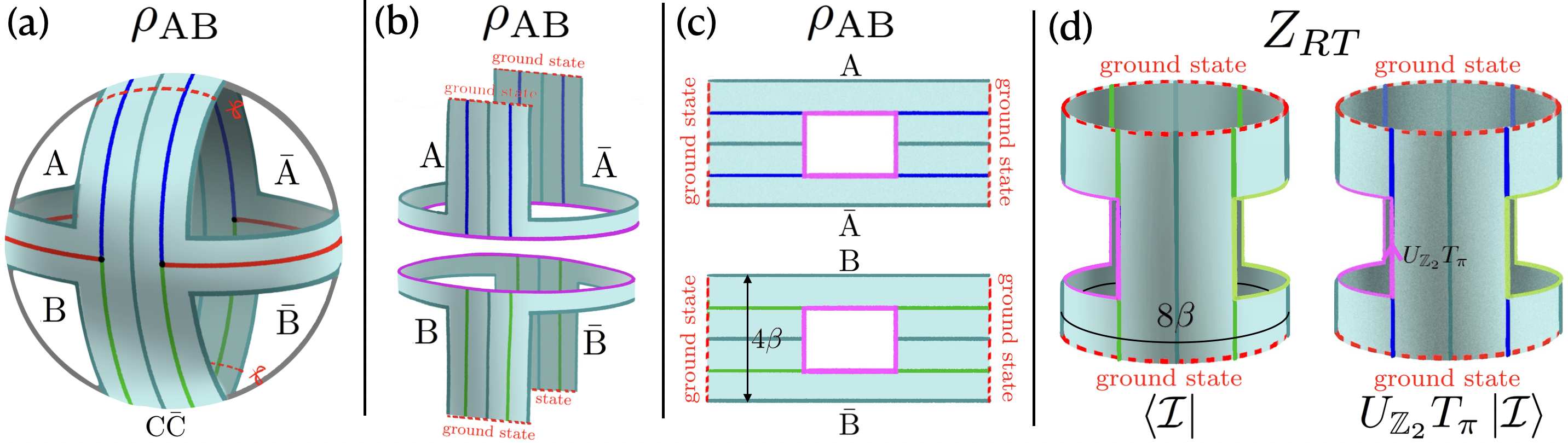}
\caption{(a): $\rho_{\text{AB}}$ is constructed by gluing a 3-vertex state and its inverse. (b) One can cut the top and bottom of $\rho_{\text{AB}}$ by inserting the projector onto a ground state. The width of the ground state is $4\beta$. We also fictitiously cut along the equator.
(c) $\rho_{\text{AB}}$ is topologically equivalent to a pair of annuli glued along the inner Boundary.(d) One can express the surface for $Z_{RT}$ as a pair of cylinders with two punctures, which are glued along each pair of punctures of the same color. One of the punctures is acted by $U_{\mathbb{Z}_2}T_{\pi}$ when gluing.}
\label{fig:constructishibashi}
\end{figure}

The reduced density matrix $\rho_{\mathrm{AB}}$ is obtained by gluing two vertex states (pair of pants) along the C boundary, so corresponds to a sphere with four punctures (see Figure \ref{fig:constructishibashi} (a)).
In the limit of $\beta\to 0$, the width of the slab is thin and only the lowest energy state propagates along the slab, where we take the spatial direction of the slab as the fictitious time direction. Therefore, one can cut the punctured sphere for $\rho_{\mathrm{AB}}$ along the slit, by inserting a projector onto a ground state cutting the thin slab with length $4\beta$ (see a red dashed line in Figure \ref{fig:constructishibashi} (a)).  

It is convenient to express $\rho_{\mathrm{AB}}$ after inserting the ground state as a pair of annuli by fictitiously cutting the surface through the equator separating the A and B hemisphere (see Figure \ref{fig:constructishibashi} (b),(c)). As outlined in the main text, when one performs partial $RT$ transformation, one needs to glue two annuli along the inner boundary associated with $\pi$ rotation. Then, the surface for $Z_{RT}$ can be obtained by preparing two copies of $\rho_{\mathrm{AB}}$ and then gluing them along the boundary.
The resulting surface is expressed as pair of cylinders with two punctures. The cylinders are glued along each puncture, one of which is associated with $\pi$ rotation due to the partial $RT$ transformation (see Figure \ref{fig:constructishibashi} (d)).

Taking $\beta\to 0$ for the punctured cylinder in Figure \ref{fig:constructishibashi} (d), each cylinder defines a state proportional to the Ishibashi state $\ket{\mathcal{I}}$ on a circle, where each puncture carries holomorphic (resp.~anti-holomorphic) Hilbert space.
To see this, let us first fold the cylinder in Figure \ref{fig:constructishibashi} (d) along the vertical line on the front and back, so that a resulting surface is two layers of thin punctured rectangles with the width $4\beta$, where each rectangle looks like the one shown in Figure \ref{fig:constructishibashi} (c).
A pair of punctures on each layer is placed at the same position after folding. The puncture at the top layer supports a holomorphic Hilbert space, while the one in the bottom supports the anti-holomorphic one since the direction of the surface is reversed upon folding.
Now, the path integral on the folded cylinder specifies a state of the non-chiral CFT on a single puncture.
Let us introduce a spatial coordinate $x$ along the boundary of the puncture, and write the stress tensor $T(x)$, $\overline{T}(x)$ for the holomorphic/anti-holomorphic Hilbert space. Noting that $T$ and $\overline{T}$ are identified at the folded boundary, in the limit of $\beta\to 0$ we have $T(x) - \overline{T}(x) = 0$ at the state. This implies that the state at $\beta\to 0$ realizes the boundary state of CFT, which is described by the Ishibashi state $\ket{\mathcal{I}}$ on the circle. 
In our case, this state $\ket{\mathcal{I}}$ is normalized as
\begin{align}
    \langle\mathcal{I}|\mathcal{I}\rangle =
	 {\rm Tr}_{\mathrm{AB}} (\rho_{\mathrm{AB}}^2),
\end{align}
which is the 2nd Renyi entropy computed in \cite{liu2023multipartite} and becomes a real quantity. Since we are mainly interested in the phase factor of $Z_{RT}$, the detail of the normalization factor is not discussed here.

$Z_{RT}$ then corresponds to the expectation value of the $\pi$ rotation operator associated with internal $\Z_2$ symmetry $U_{\Z_2}T_{\pi}$ acting on the holomorphic sector evaluated at the Ishibashi state,
\begin{align}
    Z_{RT} = \langle\mathcal{I}|U_{\Z_2}T_{\pi}|\mathcal{I}\rangle
\end{align}
where $T_\pi$ is the translation operator $T_\pi = e^{\pi i L_0}$.

To compute this, we identify the Ishibashi state as the path integral on a single annuli with thin width $\beta'\to 0$, which is regarded as another way of regularizing the Ishibashi state $\mathcal{I}$. Assuming that the invariant $I_{RT}$ does not depend on the choice of the regularization, up to overall normalization
the above expectation value is given by the torus partition function at $\beta'\to 0$ limit with $\pi$ twist (half Dehn twist) inserted,
\begin{align}
    {Z}_{RT} \propto e^{\frac{2\pi i}{48}c_-}\chi_1\left(\tau= i\beta' + \frac{1}{2},(0,1), (\mathrm{AP,AP})\right),
\end{align}
where $\propto$ denotes up to overall (real) normalization, and $(0,1)$ denotes $\Z_2$ twisted sector on a torus. 
The spin structure in both directions are anti-periodic.
The CFT character $\chi_a$ is defined as
\begin{align}
\begin{split}
 \chi_a(\tau;(0,1))&=\mathrm{Tr}_{a}[e^{2\pi i \tau (L_0-\frac{c_-}{24})} U_{\Z_2}]
    \end{split}
\end{align}

Below, we evaluate $Z_{RT}$ for a number of examples.
\subsection{$p+ip$ superconductor}

When the bulk is $n$ copies of $p+ip$ superconductors, the edge CFT is described by the $\mathrm{Spin}(n)_1$ WZW model.
 Its chiral primary fields are labeled by $\{1,\psi,\sigma\}$ when $n$ is odd, with spins $h_1=0, h_{\psi}=1/2, h_{\sigma}=n/16$.
For even $n$, the chiral primaries are labeled by $\{1,\psi,v,v'\}$
with spins $h_1=0, h_{\psi}=1/2, h_{v}=h_{v'}=n/16$.
The value of $Z_{RT}$ depends on two possible choices of $U_{\Z_2}$ that encodes the action of $RT$ symmetry; $U_{\Z_2}=1$ or $U_{\Z_2}=(-1)^F$.
\begin{itemize}
    \item When $U_{\Z_2}=1$, we have
    \begin{align}
    Z_{RT} \propto e^{\frac{2\pi i}{48}c_-}\times \left[\chi_1\left(\tau= i\beta' + \frac{1}{2}\right) +\chi_\psi\left(\tau= i\beta' + \frac{1}{2}\right)\right].
\end{align}
Each CFT character is evaluated by performing $ST^{-2}S$ transformation,
\begin{align}
\begin{split}
\chi_a\left(i\beta'+1/2\right)&=\sum_b S_{ab}\chi_b\left(-\frac{1}{i\beta'+1/2}\right) \\
&= \sum_b(ST^{-2})_{ab}\chi_b\left(\frac{2i\beta'}{i\beta'+1/2}\right)\\
&= \sum_{b}(ST^{-2}S)_{ab}\chi_b\left(\frac{i}{4\beta'}-\frac{1}{2}\right) 
\end{split}
\end{align}
and the approximation
\begin{align}
\begin{split}
    \chi_b\left(\frac{i}{4\beta'}-\frac{1}{2}\right) &\approx e^{-\pi i(h_b-\frac{c_-}{24})} e^{-\frac{2\pi}{4\beta'}(h_b-\frac{c_-}{24})}~.
    \end{split}
\end{align}
We then have
\begin{align}
\begin{split}
    Z_{RT} &\propto e^{\frac{2\pi i}{24}c_-} [(ST^{-2}S)_{11}+(ST^{-2}S)_{\psi1}] \\
    &= e^{\frac{2\pi i}{24}3c_-}\sum_a[S_{1a}+S_{\psi a}]\theta_a^{-2}d_a \\
    &\propto e^{\frac{2\pi i}{16}n}.
    \end{split}
    \end{align}

 \item When $U_{\Z_2}=(-1)^F$, we have
    \begin{align}
    \begin{split}
    Z_{RT} &\propto e^{\frac{2\pi i}{48}c_-}\times \left[\chi_1\left(\tau= i\beta' + \frac{1}{2}\right) -\chi_\psi\left(\tau= i\beta' + \frac{1}{2}\right)\right] \\
    &\propto e^{\frac{2\pi i}{24}c_-} [(ST^{-2}S)_{11}-(ST^{-2}S)_{\psi1}] \\
    &= e^{\frac{2\pi i}{24}3c_-}\sum_a[S_{1a}-S_{\psi a}]\theta_a^{-2}d_a \\
    &\propto e^{-\frac{2\pi i}{16}n}.
    \end{split}
\end{align}

\end{itemize}

\subsection{Fermionic SPT phase generating $\Z_8$ classification}
Fermionic invertible phase with unitary $\Z_2$ symmetry in (2+1)D is classified by $\Z_8\times \Z$. The $\Z_8$ is generated by a phase equivalent to $(p+ip)\times (p-ip)$ superconductor, where the $\Z_2$ symmetry acts as fermion parity on a single layer $p-ip$. In that phase, each layer contributes to $Z_{RT}$ by $e^{2\pi i/16}$, so we get $Z_{RT}=e^{2\pi i/8}$. More generally, in the fermionic invertible phase labeled by $(\nu, 2c_-)\in \Z_8\times \Z$, the partial $RT$ evaluates as
\begin{align}
    Z_{RT} \propto e^{\frac{2\pi i}{8}\nu }e^{\frac{2\pi i}{16}2c_-}.
\end{align}
\subsection{Chern insulator with $RT$ symmetry}
The Chern insulator can have a $RT$ symmetry which conjugates the $U(1)^f$ charge. In this setup, the Chern insulator is equivalent to the 2 copies of $p+ip$ superconductors, and $U(1)^f$ acts by continuous symmetry that acts on the layer index. $RT$ symmetry again acts on the fermion parity symmetry in the single layer $p+ip$. In this case, one layer of $p+ip$ contributes to $Z_{RT}$ by $e^{2\pi i/16}$, while the other by $e^{-2\pi i/16}$. So, the Chern insulator has a trivial $Z_{RT}$,
\begin{align}
    Z_{RT} \propto 1.
\end{align}

\subsection{Bosonic SPT phase generating $\Z_2$ classification}
Finally, let us consider the setup where the bulk is the bosonic SPT phase with $RT$ symmetry in (2+1)D. Since $c_-=0$, $Z_{RT}$ is expressed as
\begin{align}
    Z_{RT} \propto  \chi_1\left(\tau= i\beta' + \frac{1}{2},(0,1)\right)\propto S_{(0,1)(1,0)}\theta_{(1)}^2S_{(1,0)(0,1)},
\end{align}
where $S_{(g,h)(h,g)}$ with $g,h\in \Z_2$ is the $S$ matrix of the characters with $\Z_2$ defect insertions, and $\theta_{(1)}=e^{2\pi i h_{(1)}}$ is the spin carried by the vacuum of the $\Z_2$ twisted sector, which satisfies $h_{(1)}= 1/4$ mod 1/2. 
This implies that
\begin{align}
    Z_{RT} \propto -1.
\end{align}

\section{Classification for $(2+1)D$ invertible fermionic topological phases with ${\bf R}{\bf T}$ symmetry}
\label{app:classification}

\subsection{Generalities of fermionic symmetry groups}
Here we introduce generic terminologies about fermionic symmetry groups. In general, the (0-form) symmetry group of a fermionic system $G_f$ has the structure of the central extension of a bosonic symmetry group $G_b$ by ${\mathbb Z}_2^f$,
\begin{equation}
    1 \rightarrow {\mathbb Z}_2^f \rightarrow G_f \rightarrow G_b \rightarrow 1. 
\end{equation}
This central extension is specified by a $2$-cocycle $\omega_2 \in {\mathcal H}^2(G_b, {\mathbb Z}_2)$. More specifically, one can denote a general element in $G_f$ as a pair $({\bm g}, a)$, where ${\bm g} \in G_b$ and $a \in {\mathbb Z}_2^f = \{0, 1 \}$, then the group law in $G_f$ reads
\begin{equation}
    ({\bf g}_1, a_1) ({\bf g}_2,a_2) = ({\bf g}_1{\bf g}_2, a_1 + a_2 + \omega_2({\bf g}_1,{\bf g}_2)).
\end{equation}

We also define the homomorphism $s_1: G_b \rightarrow {\mathbb Z}_2$. If ${\bf g}$ is unitary, $s_1({\bf g}) =0$, if ${\bf g}$ is antiunitary, $s_1({\bf g}) = 1$.

\subsection{Crystalline equivalence principle}\label{Sec_fCEP}

In general, for a clean 2D system without any defects defined on the infinite plane, the spatial elements in $G_b$ are specified by a map $(\vec R, \rho_s): G_b \rightarrow {\mathbb R}^2 \rtimes {O}(2)$ where ${\mathbb R}^2 \rtimes {O}(2)$ is the group of continuous translations (${\mathbb R}^2$), rotations (${SO}(2)$ part of ${O}(2)$), and reflections (${\mathbb Z}_2$ part of ${O}(2)$) in two dimensions. Note that, the map $\vec R: G_b \rightarrow {\mathbb R}^2$, and $\rho_s : G_b \rightarrow {O}(2)$. Having a non-trivial map $(\vec R, \rho_s)$ implies that $G_b$ acts as a spatial symmetry, and we include it as part of the symmetry data of the system. 

The fermionic crystalline equivalence principle (fCEP) states that the classification of invertible fermionic topological phases with spatial symmetry $G_f$ defined by the data $(G_b, s_1, \omega_2, (\vec R,\rho_s))$ is one-to-one correspondence with the classification of invertible fermionic topological phases with an effective internal symmetry $G_f^{\rm eff}$ that has data $(G_b, s_1^{\rm eff}, {\omega_2}^{\rm eff})$ where $G_b$ acts trivially on space. The terms $s_1^{\rm eff}$, $\omega_2^{\rm eff}$ are conjectured to be determined by $s_1,\omega_2,\rho_s$ through the equations~\cite{Manjunath_2023_PRB_Nonperturbative} 
\begin{subequations}
    \begin{align}
        s_1^{\rm eff} &= s_1 + w_1, \label{S1Effective}\\
        \omega_2^{\rm eff} &= \omega_2 + w_2 + w_1(s_1 + w_1), \label{Omega2Effective}
    \end{align}
\end{subequations}
where $w_1 = \rho_s^*{\rm w}_{1,r}$, $w_2 = \rho_s^* {\rm w}_2$ are obtained by pulling back the Stiefel-Whitney classes ${\rm w}_{1,r}$, ${\rm w}_2$ that generate ${\mathcal H}^1({O}(2),{\mathbb Z}_2)$, ${\mathcal H}^2({SO}(2), {\mathbb Z}_2)$.

Let us consider the examples $G_b = {U}(1) \rtimes {\mathbb Z}_2^{RT}, {U}(1) \times {\mathbb Z}_2^{RT}, \Z_2^{RT}$. In these cases, $RT$ can be viewed as an effective anti-unitary reflection symmetry, which implies that $s_1, w_1$ are both non-trivial. Since there is no effective spatial rotation in $G_b$, 
we have $w_2 =0$.   
Thus we have the effective data:
\begin{equation}
    \begin{aligned}
        s_1^{\rm eff} &= s_1 + w_1 = 0, \\
        \omega_2^{\rm eff} &= \omega_2 + w_2 + w_1(s_1 + w_1) = \omega_2.
    \end{aligned}
\end{equation}
The above result $s_1^{\rm eff} = 0, \omega_2^{\rm eff}=\omega_2$ implies that $\Z_2^{RT}$ is treated as a unitary internal $\Z_2$ symmetry under fCEP, and internal symmetry $G_f^{\rm eff}$ has the same group structure as $G_f$.

\subsection{Classifications}
\subsubsection{General rules}

After converting the data from crystalline symmetry $G_f$ to internal symmetry $G_f^{\rm eff}$, the data for classification of invertible fermion phases are~\cite{Barkeshli_2022_PRB_Classification} $(c_-, n_1, n_2, \nu_3) \in \frac{1}{2} {\mathbb Z} \times C^1(G_b, {\mathbb Z}_2) \times C^2(G_b, {\mathbb Z}_2) \times C^3(G_b, U(1)_T)$. These data satisfies the following equations:
\begin{subequations}
    \begin{align}
        dn_1 &= 0 \quad {\rm mod}~2, \label{Eqn1}\\
        dn_2 &= n_1 \cup(\omega_2^{\rm eff} + s_1^{\rm eff} \cup n_1) + c_- \omega_2^{\rm eff} \cup_1 \omega_2^{\rm eff} \quad {\rm mod}~2, \label{Eqn2}\\
        d\nu_3 &= {\mathcal O}_4[c_-,n_1,n_2], \label{Eqnu3}
    \end{align}
\end{subequations}
where the form of ${\mathcal O}_4$ will be given later. Below, we explicitly compute the classification with $G_f = U(1)^f\rtimes \Z_2^{RT},U(1)^f\times \Z_2^{RT}$.
The case with $G_f^{\text{eff}} =\Z_2^f\times \Z_2^{RT}$ is treated as $G_f = \Z_2^f \times \Z_2$ and has been computed in previous works, e.g., Ref.~\cite{Barkeshli_2022_PRB_Classification,gu2014}

\subsubsection{$G_f =U(1)^f\rtimes \Z_2^{RT}$}
Here let us compute the classification for $G_f =U(1)^f\rtimes \Z_2^{RT}$ which is equivalent to the internal symmetry $\mathrm{O}(2)^f$ under fCEP, with $\omega_2=w_2$ (in this section, $w_1,w_2$ means $w_1(O(2)), w_2(O(2))$). 
Note that $c_-$ has to be integer since we have $U(1)^f$ symmetry.
Here let us set some background for computing the consistency equations:
\begin{itemize}
\item The element of O(2) is written as a pair $({\bf g}, \sigma)\in U(1)\times \Z_2$. 
The O(2) gauge field is then expressed as assignment of $({\bf g},\sigma)$ on each 1-simplex, together with the lift of $U(1)$ field ${\bf g}_{01}$ to $\mathbb{R}$ on each 1-simplex, which is denoted as $\hat{\bf g}_{01}$.
The field strength of $U(1)$ gauge field can be expressed as a (twisted) coboundary of $\hat{\bf g}$ given by $F(012) =\frac{\hat{\bf g}_{01} +^{\sigma_{01}}\hat{\bf g}_{12} -\hat{\bf g}_{02} }{2\pi}$. Here $^{\sigma}\hat{\bf g}$ denotes the action of the charge conjugation element $\sigma\in \Z_2$ which acts on $\hat{\bf g}$ by $\hat{\bf g} \to (-1)^\sigma\hat{\bf g}$.

\item The Stiefel-Whitney classes are expressed at the cochain level by $w_2(012) =\frac{\hat{\bf g}_{01} +^{\sigma_{01}}\hat{\bf g}_{12} -\hat{\bf g}_{02} }{2\pi}$, $w_1(01) = \sigma(01)$ mod 2.

\item It is convenient to set up terminology of twisted cohomology theory. The twisted cohomology is denoted as $H_{\rho}^*(M,X)$, where $X$ is an Abelian group with $G$ action $\rho: G\to\mathrm{Aut}(X)$. For a given configuration of the $G$ gauge field $g\in Z^1(M,G)$, the twisted coboundary is defined for $\omega\in C^d(M,X)$ as
\begin{align}
    d_{\rho}\omega_{(01\dots d+1)}=\rho_{g_{1,0}}[\omega_{(1\dots d+1)}] +    \sum_{i=1}^{d+1}(-1)^{i}\cdot \omega_{(0\dots\hat{i}\dots d+1)},
\end{align}
where $\hat{i}$ means skipping over $i$.
One can see that $d_{\rho}d_{\rho}=0$, so it defines a cohomology twisted by the $G$ action $H_{\rho}^d(M,X):=Z_{\rho}^d(M,X)/B_{\rho}^d(M,X)$.
We also use a twisted version of cup product
\begin{align}
    -\cup_{\rho}-:C^k(M,\Z_2)\times C^l(M,\Z_2)\to C^{k+l}(M,\Z_2),
\end{align}
whose explicit form is written as
\begin{align}
    (\alpha\cup_\rho\beta)_{(0,\dots, k+l)} = \alpha_{(0,\dots,k)}\cdot\rho_{g_{k,0}}[\beta_{(k,\dots,k+l)}].
    \label{eq:cupdef}
\end{align}
The cup product satisfies the twisted Leibniz rule at the cochain level,
\begin{align}
    d_{\rho}(\alpha\cup_\rho\beta)=(d_{\rho}\alpha)\cup_\rho\beta + (-1)^k\cdot \alpha\cup_\rho(d_{\rho}\beta).
\end{align}
In our case, we want to take $G=\mathbb{R}$, and $\rho$ is charge conjugation action $\sigma$ on $\mathbb{R}$. Then, we have $w_2(012) = d_\sigma\hat{\bf g}(012)$.

\item One can check $w_1w_2 = \frac{d w_2}{2}$. This can be seen by noticing $d_\sigma w_2=0$, and
\begin{align}
    d w_2(0123) = d_\sigma w_2(0123) - ^{\sigma_{01}}w_2(123) + w_2(123) = (1 + (-1)^{\sigma(01)})w_2(123) = 2\sigma(01)w_2(123).
\end{align}
So we have $d w_2 = 2w_1w_2$, hence $w_1w_2 = \frac{d w_2}{2}$.
\end{itemize}
Below, let us solve the consistency equations and derive the classification of invertible phases.

\paragraph{Solution for the $d n_2$ equation:}

For the equation of $d n_2$, noting that $s_1^{\mathrm{eff}}=0, \omega_2^{\mathrm{eff}}=w_2$ we have
\begin{equation}\label{Eq_n2}
    d n_2 = n_1 \cup w_2 + c_- w_2\cup_1 w_2 \quad ({\rm mod}~2).
\end{equation}
We have $w_2\cup_1 w_2 = \mathrm{Sq}^1w_2 = d w_2/2 = w_1w_2$. We then have $d n_2 = (n_1+c_- w_1)w_2$, from which we must have 
\begin{align}
    n_1 = c_-w_1 \quad \mod 2,
\end{align}
in order to find a solution. This constraint has been obtained in \cite{Manjunath_2023_PRB_Nonperturbative}. When this condition is satisfied, $n_2$ generally takes its value in ${\mathcal H}^2({O}(2), {\mathbb Z}_2)$.

\paragraph{Solution for the $d\nu_3$ equation with even $c_-$:}
When $c_-$ is even, let us write $c_-=2k$ with $k\in\Z$. The obstruction class $\mathcal{O}_4$ with $d\nu_3 = \mathcal{O}_4$ is given by
    \begin{align}
        \mathcal{O}_4 = (-1)^{n_2(n_2+w_2)} e^{\frac{2\pi ik}{4}\mathcal{P}(w_2)}. 
    \end{align}
Pontryagin square is rewritten as
\begin{align}
    \mathcal{P}(w_2) = w_2w_2- w_2\cup_1d w_2 = w_2w_2 - 2w_2 \cup_1 (w_1w_2) = w_2w_2 + 2 (w_1\cup_1 w_2)w_2 + 2w_1^2w_2 + 2d( w_2\cup_2 (w_1w_2))\quad \text{mod 4}.
\end{align}
So one can rewrite the equation as
\begin{align}
    \mathcal{O}_4 = (-1)^{n_2(n_2+w_2) + kw_1^2w_2} e^{\frac{2\pi i k}{4} (w_2 +2(w_1\cup_1 w_2))w_2} (-1)^{d(k w_2\cup_2 (w_1w_2))}
\end{align}
One can check that 
$(w_2 +2(w_1\cup_1 w_2))w_2(01234) = (1+2w_1(02))w_2(012)w_2(234) = w_2\cup_\sigma w_2 (01234)$. So
\begin{align}
    (w_2 +2(w_1\cup_1 w_2))w_2 = w_2\cup_\sigma w_2 = d_\sigma (\hat{\bf g}\cup w_2 ) = d(\hat{\bf g}\cup w_2). 
\end{align}
In the last equation, we used that $\hat{\bf g}\cup w_2$ is even under charge conjugation, so twisted and untwisted coboundary is the same. So
\begin{align}
    \mathcal{O}_4 = (-1)^{n_2(n_2+w_2) + kw_1^2w_2} e^{\frac{2\pi i k}{4} d(\hat{\bf g}\cup w_2)} (-1)^{d(k w_2\cup_2 (w_1w_2))}
\end{align}
Then, this $\mathcal{O}_4$ has a trivialization $d\nu_3 = \mathcal{O}_4$ only if $n_2 = kw_1^2, kw_1^2 + w_2$. These two data actually specifies the same phase due to the equivalence relations \cite{Barkeshli_2022_PRB_Classification}. Then we can find a solution
\begin{align}
    \nu^{(0)}_3 = e^{\frac{2\pi i k}{4}(w_1^3+\hat{\bf g}\cup w_2)} (-1)^{k w_2\cup_2 (w_1w_2)}
\end{align}
Summarizing, we have a solution with fixed even $c_-=2k$, given by
\begin{align}
    (n_1,n_2,\nu_3) = \left(0, kw_1^2,\nu_3^{(0)} + \nu_3'\right),
\end{align}
where $\nu'_3\in \mathcal{H}^3(O(2), U(1))$.

\paragraph{Classification of invertible phases with generic $c_-$:}
Now let us consider the case where $c_-$ is a generic integer. In that case we have $n_1=c_-w_1$, implying that the defect of the $\Z_2^{RT}$ symmetry bounds a Majorana zero mode when $c_-$ is odd~\cite{Manjunath_2023_PRB_Nonperturbative}. 
While we do not try to provide complete solutions to the consistency equations $(n_2,\nu_3)$ when $c_-$ is odd, one can still find the classification of the invertible phases. Since the Chern insulator is consistent with $\Z_2^{RT}$ symmetry conjugating $U(1)^f$, there is one non-trivial phase carrying $c_-=1$. 
Meanwhile, the previous analysis with $c_-=0$ implies that all the non-trivial invertible phase with $c_-=0$ are the bosonic SPT phases classified by $\mathcal{H}^3(O(2), U(1)) = \Z\times \Z_2$. This implies that the classification is generated by the Chern insulator $\Z$ and bosonic SPT phases $\Z\times \Z_2$, which together form the whole classification $\Z^2\times \Z_2$. 
The invertible phases are labeled by $(c_-, C, \nu)$ with $\nu\in\Z_2$ the index characterizing the $\Z_2$ classification of the bosonic SPT.

\subsubsection{$G_f = U(1)^f \times \Z_2^{RT}$}
Here let us compute the classification for $G_f =U(1)^f\times \Z_2^{RT}$ which is equivalent to the internal symmetry $U(1)^f\times \Z_2$ under fCEP, with $\omega_2=w_2$ (in this section, $w_1$ is a non-trivial element in $\mathcal{H}^1(\Z_2,\Z_2)$, and $w_2=w_2(U(1))$). Note that $c_-$ has to be integer since we have $U(1)^f$ symmetry. Below, let us derive the classification of invertible phases by computing the consistency conditions.

First of all, we have
\begin{equation}
    dn_2 = n_1 \omega_2 + c_- \omega_2 \cup_1 \omega_2.
\end{equation}
Note that $\omega_2 \cup_1 \omega_2 = \mathrm{Sq}^1 w_2 = 0$, so we get $dn_2 = n_1 \omega_2$. In order to have a solution, we must have $n_1 = 0$. $n_2$ is then closed and $n_2 \in {\mathcal H}^2 ({U}(1) \times \Z_2, \Z_2)$, which take the value of $a w_1^2 + b\omega_2$, $a,b \in \{0,1\}$. Here, $b=0,1$ actually corresponds to the redundant labels for the invertible phase due to the equivalence relations \cite{Barkeshli_2022_PRB_Classification}, so we only need to consider the case $n_2=0$ or $n_2=w_1^2$ below. 

Since $n_1 = 0$, the $\mathcal{O}_4$ class is given by
\begin{equation}
    {\mathcal O}_4 = (-1)^{n_2 \cup (n_2 + \omega_2)} e^{i \pi \frac{c_-}{4} {\mathcal P}(\omega_2)}.
\end{equation}
Let us first consider the case $c_-=0$ and find solutions for $d\nu_3=\mathcal{O}_4$.

When $n_2 = 0$, we have $d\nu_3=0$ and $\nu_3 \in {\mathcal H}^3({U}(1)\times \Z_2, {U}(1)) = \Z \times \Z_2^2$ labels the bosonic invertible phases, with the $\Z$ from the integer quantum Hall effect (IQHE), one $\Z_2$ generated by $w_1^3/2$, and the other $\Z_2$ generated by $w_1\omega_2/2$. However, since for $(n_2,\nu_3)$ we have the equivalence of phases $(0,0)\sim (db,b\omega_2/2)$, by taking $b = w_1$, one can see that $(n_2,\nu_3)=(0,w_1\omega_2/2)$ is a trivial phase~\cite{Barkeshli_2022_PRB_Classification}. This implies that the phases with $c_-=0, n_2=0$ is labeled by $\Z\times \Z_2$.

When $n_2 = w_1^2$, since
\begin{equation}
   \frac{1}{2} n_2(n_2 + \omega_2) = \frac{1}{2}w_1^2 (w_1^2 + \omega_2) = \frac{dw_1}{4}\big{(} \frac{dw_1}{2} + \omega_2 \big{)} = \frac{1}{4}d\big{(} w_1\frac{dw_1}{2} +w_1 \omega_2 \big{)},
\end{equation}
then we have the solution of $\nu_3$ as
\begin{align}
    \nu_3 = \frac{1}{4}\big{(} w_1\frac{dw_1}{2} +w_1 \omega_2 \big{)} + \nu_3'
\end{align}
with $\nu'_3 \in {\mathcal H}^3({U}(1)\times \Z_2, {U}(1))$. 

To obtain the classification of invertible phases with $c_-=0$, we further notice that the stacking rule of the phases with the above $(n_2,\nu_3)$ can be non-trivial. Indeed, the stacking rule of the phases with $n_2=w_1^2$ is found to be
\begin{align}
    (w_1^2,\frac{1}{4}\big{(} w_1\frac{dw_1}{2} +w_1 \omega_2 \big{)}) \times (w_1^2,\frac{1}{4}\big{(} w_1\frac{dw_1}{2} +w_1 \omega_2 \big{)})
    = (0, \frac{1}{2} w_1^3),
\end{align}
which implies that two copies of the fermionic SPT phases with $n_2=w_1^2$ is equivalent to a bosonic SPT phase. The generic formula for the stacking rule of fermionic invertible phases is found in \cite{Barkeshli_2022_PRB_Classification}.
This implies that for the phases with $n_2=0$ labeled by $\Z\times\Z_2$, the $\Z_2$ classification gets extended by the fermionic phase with $n_2=w_1^2$.
Therefore, the classification with $c_-=0$ is given by $\Z\times \Z_4$.

Additionally, we have the phase with $c_-=1$ on the top of the $c_-=0$ classification, so the whole classification of invertible phases is given by $\Z^2 \times \Z_4$, we have one $\Z$ from $c_-$, one $\Z$ from bosonic IQHE, and the other $\Z_4$ comes from a fermionic SPT phase.

\section{Numerical methods}
\subsection{Standard numerical procedure for entanglement Hamiltonian}

In this section we discuss how to derive the entanglement Hamiltonian and compute the RT invariant from a free-fermion tight-binding Hamiltonian based on Ref.~\cite{Shiozaki2018antiunitary,cheong2003manybody,Borchmann2014Entanglement,Peschel_2003,Peschel_2009_JPA_Reduced, Shapourian2017Partial}. Consider the tight-binding Hamiltonian given by:
\begin{equation}
	H = \sum_{<ij>} -t_{ij} f_i^\dagger f_j + \Delta_{ij} f_i^\dagger f_j^\dagger + {\rm H.c.}. 
\end{equation} 
where the fermionic operators satisfy the anticommutation relation: $\{ f_i, f_j^\dagger \} = \delta_{ij}$.  The reduced density matrix of such Hamiltonians in the subsystem $\text{A}\cup\text{B}$ of the whole system $\text{E} = {\rm A} \cup {\rm B} \cup {\rm C}$ can be expressed in terms of a quadratic form:
\begin{equation}\label{Eq_Reduced_Density_Matrix_Definition}
    \rho_{\text{AB}} = {\rm Tr}_{\rm{C}}(\rho) = \frac{e^{-\hat H_{\text{AB}}}}{\mathcal Z},
\end{equation}
where the entanglement Hamiltonian is defined as:
\begin{equation}
    \hat H_{\text{AB}} = \sum_{i,j} h_{ij}^1 f_i^\dagger f_j + h^2_{ij} f_i^\dagger f_j^\dagger + {\rm H.c.},
\end{equation}
and ${\mathcal Z}$ is the normalization factor. We denote the number of the degrees of freedom (the product of the number of sites and the number of internal degrees of freedom per site) as $N_{\text{AB}}$. 

The Hamiltonian described above can be expressed in the Nambu basis as: 
\begin{equation}
	\hat H_{\text{AB}} = \psi^\dagger {\mathcal H} \psi,
\end{equation}
where $\psi = (f_1, ..., f_{N_{\text{AB}}}, f^\dagger_1, ... f^\dagger_{N_{\text{AB}}})^T$. The matrix ${\mathcal H}$ exhibits particle-hole symmetry, enabling it to be diagonalized as ${\mathcal H} = WDW^\dagger$, where $D = {\rm diag} (E, -E)$, $E = {\rm diag}(E_1,\cdots, E_{N_{\text{AB}}})$ with $E_i >0$, $\forall i$. $W$ is the eigenvectors.
We can further define the correlation matrix $G$ as:
\begin{equation}
	G = \begin{pmatrix}
		\langle f_if_j^\dagger \rangle & \langle f_i f_j \rangle \\
		\langle f_i^\dagger f_j^\dagger \rangle & \langle f_i^\dagger f_j \rangle
	\end{pmatrix}=\begin{pmatrix}
		[1-C^T]_{ij}& [F^{\dagger}]_{ij}  \\
		   F_{ij}  & C_{ij} 
	\end{pmatrix}
\end{equation}
One can demonstrate that $G$ can be represented as $G = W G_c W^\dagger$, where $G_c = {\rm diag}(1-{\mathcal F},{\mathcal F})$, with ${\mathcal F} = {\rm diag}(n_f(E_1)...n_f(E_N))$ with $n_f(x) = 1/(1 + e^{x/(k_BT)})$. In our numerics, we take the 0 temperature limit which fills every negative energy state.

The reduced density matrix can be formulated in the coherent state basis as follows:
\begin{equation}\label{Eq_DM_Coherent_State}
    \rho_{\text{AB}} = \frac{1}{{\mathcal Z}_\rho} \int d[\xi] d[\bar \xi] \exp \bigg{[} \frac{1}{2} \sum_{i,j\in I} (\xi_i, \bar{\xi}_i) S_{ij}
(\xi_j, \bar{\xi}_j)^T  \bigg{]} \ket{\{ \xi_j \}_{j \in \text{AB}}} \bra{\{ \bar \xi_j \}_{j \in \text{AB}}},
\end{equation}
where $(\xi_j, \bar{\xi}_j)^T$ denotes the column vector of $(\xi_j, \bar{\xi}_j)$. The $S_{ij}$ matrix is given by: 
\begin{equation}\label{eq:gammaToS}
    S_{ij} = \Gamma_{ij} + [i\sigma_y \otimes\mathbb{I}_{N_{\text{AB}} \times N_{\text{AB}}}]_{ij}. 
\end{equation}
where $\sigma_y = \left( \begin{matrix} 0 & -i \\ i & 0 \end{matrix} \right)$ is the second Pauli matrix that mixes $\xi_i$ and $\bar \xi_i$. The inverse of the matrix $\Gamma$ can be formulated using correlators:
\begin{equation}
    [\Gamma^{-1}]_{ij} = \begin{pmatrix}
        [F^\dagger]_{ij} & [C^T]_{ij} \\
        -C_{ij} & F_{ij}
    \end{pmatrix},
\end{equation}
which is a $2N_{\text{AB}} \times 2N_{\text{AB}}$ matrix.
The normalization factor ${\mathcal Z}_\rho$ can be evaluated as ${\mathcal Z}_\rho = {\rm Pf}[\Gamma]$.

\subsection{Partial time reversal}
We now proceed to calculate the partial time-reversal transformation of $\rho_{\text{AB}}$. 
Consider the case where the time reversal $T=\mathcal{K}$. In this case, the partial time reversal of  $\rho_{\text{AB}}$ is equivalent to the partial transpose $\rho_{\text{AB}}^{T_{\text{A}}}$.
We begin by defining $\rho_{\text{AB}}^{T_{\text{A}}}$ while specifying which region each site belongs to: 
\begin{equation}
    \rho_{\text{AB}}^{T_{\text{A}}} = \frac{1}{{\mathcal Z}_\rho} \int d[\xi] d[\bar \xi] \exp \bigg{[} \frac{1}{2} \sum_{i,j \in \text{AB}} (\xi_i, \bar{\xi}_i) S_{ij}
(\xi_j, \bar{\xi}_j)^T   \bigg{]} \times \ket{\{ i \bar \xi_j \}_{j \in A}, \{\xi_j \}_{j \in B}}\bra{\{ i \xi_j \}_{j \in A}, \{ \bar \xi_j \}_{j \in B}}.
\end{equation}

This transformation can be incorporated through the redefinition of the $S_{ij}$ matrix after introducing the new variables ${\bf \xi} = U_S \chi$, yielding:
\begin{equation}\label{Eq_Coherent_Path_Integral}
    \rho_{\text{AB}}^{T_\text{A}} = \frac{1}{{\mathcal Z}_\rho} \int d[\chi]d[\bar \chi] \exp \bigg{[} \frac{1}{2} \sum_{i,j \in I} (\chi_i,  \bar{\chi}_i) S_{ij}^{T_\text{A}}(\chi_j,  \bar{\chi}_j)^T
       \bigg{]} \ket{\{ \chi_j\}_{j \in {\text{AB}}}}\bra{\{ \bar \chi_j\}_{j \in \text{AB}} },
\end{equation}
where $S^{T_\text{A}} := U_S^T S U_S$ and $U_S = U_S^T$ act as a permutation matrix defined by:
\begin{equation}
    U_S = \begin{pmatrix}
        0 & 0 & -i{\mathbb I}_{\text{A}} & 0 \\
        0 & {\mathbb I}_\text{B} & 0 & 0 \\
        -i{\mathbb I}_\text{A} & 0 & 0 & 0 \\
        0 & 0 & 0 & {\mathbb I}_\text{B}
    \end{pmatrix},
\end{equation}
in the $(\{\xi_j \}_{j \in \text{A}}, \{ \xi_j \}_{j \in \text{B}}, \{ \bar \xi_j \}_{j \in \text{A}}, \{ \bar \xi_j \}_{j \in \text{B}})$ basis. Here, ${\mathbb I}_\text{A}$ and ${\mathbb I}_\text{B}$ are identity matrices acting on regions A and B, respectively. 

\subsection{Partial reflection}
In a manner similar to the partial time reversal, the partial reflection can also be incorporated by redefining the $S$ matrix. We apply the combined $RT$ symmetry action on the $S$ matrix as $S^{RT_\text{A}} := U_R^T S^{T_\text{A}} U_R$, where
\begin{equation}
    U_R = \begin{pmatrix}
        \mathcal{R}_{\text{A}} & 0 & 0 & 0 \\
        0 & {\mathbb I}_{\text{B}} & 0 & 0 \\
        0 & 0 & \mathcal{R}_\text{A}^* & 0 \\
        0 & 0 & 0 & {\mathbb I}_\text{B}
    \end{pmatrix},
\end{equation}
and $\mathcal{R}_\text{A}$ denotes the reflection matrix acting within region A. For any $j$ labeled by its position $x,y$ and layer index $s$, $\mathcal{R}_\text{A}$ is defined by its action $\mathcal{R}_\text{A}\xi_{x,y,s}=\xi_{-x,y,s}$. 

\subsection{Calculation of $Z_{RT}$}
To compute the topological invariant $Z_{RT}=\text{Tr}(\Xi)$, where $\Xi := \rho_{\text{AB}} R_{\rm A}\rho_{\text{AB}}^{T_\text{A}}R_{\rm A}^{\dagger}$, we begin by multiplying the two density matrices and then performing a partial Gaussian integral as follows:
\begin{equation}
    \begin{aligned}
\Xi &= \frac{1}{{\mathcal Z}_\rho^2} \int d[\chi]d[\bar \chi] d[\xi] d[\bar \xi] e^{\frac{1}{2} \sum_{i,j \in {\text{AB}}} (\chi_i,\bar{\chi}_i) S_{ij} (\chi_j,\bar{\chi}_j)^T} e^{\frac{1}{2} \sum_{i,j \in \text{AB}} 
(\xi_i, \bar{\xi}_i) S_{ij}^{RT_{\text{A}}}
(\xi_j, \bar{\xi}_j)^T   } 
\ket{\{ \chi_j\}_{j \in \text{AB}}}\bra{\{ \bar \chi_j \}_{j \in \text{AB}}} \ket{\{ \xi_j\}_{j \in \text{AB}}} \bra{\{ \bar\xi_j \}_{j \in \text{AB}}} \\
    &= \frac{\mathscr{Q}}{{\mathcal Z}_\rho^2} \int d[\chi] d[\bar \xi] e^{\frac{1}{2}\sum_{i,j \in \text{AB}}(\chi_i, \bar \xi_i)\tilde S_{ij} (\chi_j,\bar \xi_j)^T} \ket{\{\chi_j\}_{j \in \text{AB}}}\bra{\{\bar \xi_j\}_{j \in \text{AB}}},
    \end{aligned}
\end{equation}
where $\mathscr{Q}$ is a normalization factor. The partial Gaussian integral can be further detailed as:
\begin{equation}
    \mathscr{Q}e^{\frac{1}{2}\sum_{i,j\in\text{AB}} (\chi_i,\bar \xi_i) \tilde S_{ij}(\chi_j ,\bar \xi_j)^T } = \int [d\bar \chi] [d\xi] e^{\sum_{i,j \in I} (\chi_i,\bar \chi_i,\xi_i, \bar \xi_i) M_{ij} (\chi_i \bar \chi_i, \xi_i, \bar \xi_i)^T}
\end{equation}
with the kernel $M$ defined as:
\begin{equation}
    M = \frac{1}{2} \begin{pmatrix}
        S & K \\
        -K^T & S^{T_A}
    \end{pmatrix} =: \left(\begin{array}{@{}c|@{}c@{}|c@{}}

        \alpha_1 & \quad\beta_1\quad & \alpha_2 \\
        \hline
        -\beta_1^T & \gamma & -\beta_2^T \\
        \hline
        \alpha_3 & \beta_2 & \alpha_4
\end{array}\right), \quad K = \begin{pmatrix}
        0 & 0 \\
        -{\mathbb I} &0
    \end{pmatrix}.
\end{equation}
The submatrix $K$ arise from the inner product $\langle \{ \bar \chi_j \}_{j \in I} | \{ \xi_j \}_{j \in A} \rangle$. We decompose $M$ into block matrices, with $\{\alpha_1,\alpha_2,\alpha_3,\alpha_4\}$ as $N_{\text{AB}}\times N_{\text{AB}}$ matrices, $\{\beta_1,\beta_2\}$ as $2N_{\text{AB}}\times N_{\text{AB}}$ matrices, and $\gamma$ as a $2N_{\text{AB}}\times 2N_{\text{AB}}$ matrix. We also define $\alpha$ and $\beta$ from these block matrices:
\begin{equation}
    \alpha = \begin{pmatrix}
        \alpha_1 & \alpha_2 \\
        \alpha_3 & \alpha_4
    \end{pmatrix}, \quad \beta = \begin{pmatrix}
        \beta_1 \\
        \beta_2
    \end{pmatrix},
\end{equation}
where $\alpha,\beta$ and $\gamma$ are square matrices with the same sizes.

After performing the partial Gaussian integral, $\tilde S_{ij}$ can be expressed as:
\begin{equation}
    \tilde S_{ij} = [\alpha + \beta(\gamma)^{-1} \beta^T]_{ij}.
\end{equation}
We also calculate the normalization factor $\mathscr{Q}=\text{Pf}[\gamma]$. The new $\tilde{\Gamma}$ matrix is defined using Eq.~\eqref{eq:gammaToS}
\begin{equation}
    \tilde \Gamma_{ij} = \tilde S_{ij} - [i\sigma_y \otimes {\mathbb I}_{N_{\text{AB}}\times N_{\text{AB}}}]_{ij}. 
\end{equation}
Finally, we obtain the $RT$ invariant $Z_{RT}$ by performing the Gaussian integral:
\begin{equation}
    Z_{RT} = {\rm Tr}[\Xi] = \frac{{\rm Pf}[\gamma]{\rm Pf}[\tilde \Gamma]}{{\rm Pf}[\Gamma]^2},
\end{equation}
which gives the numerical value for Eq.~(1) in the main text.

\vfill

\end{document}